\documentclass[showpacs,twocolumn,aps]{revtex4}
\usepackage{amssymb}
\usepackage{amsmath}
\usepackage{graphicx}
\usepackage{lscape}
\usepackage{booktabs}

\begin{document}

\title{Pseudorapidity dependence of short-range correlations from a multi-phase transport model}

\author{%
      Wang Meijuan$^{1}$\footnote{wangmj@cug.edu.cn},%
\quad CHEN Gang$^{1}$ \quad Ma Guoliang$^{2}$
 \quad Wu Yuanfang$^{3}$ }

\address{
$^1$Physics Department, China University of Geoscience, Wuhan
430074, China
 \\
 $^2$Shanghai Institute of Applied Physics, Chinese Academy of Sciences, Shanghai 201800, China
 \\
  $^3$ Key Laboratory of Quark and Lepton Physics (MOE) and Institute of Particle
Physics, \\Central China Normal University, Wuhan 430079, China}

\begin{abstract}
 Using a multi-phase transport model (AMPT) that includes both initial partonic and hadronic interactions,
 we study neighboring bin multiplicity correlations as a function of pseudorapidity in Au+Au collisions
 at  $\sqrt{s_{NN}} = 7.7-62.4$ GeV.
 It is observed that for $\sqrt{s_{NN} } < $19.6GeV Au+Au collisions, the short-range correlations of final particles
 have a trough at central pseudorapidity,
 while for $\sqrt{s_{NN}} > $19.6GeV AuAu collisions, the short-range correlations of final particles
 have a peak at central pseudorapidity.
 Our findings indicate that the pseudorapidity dependence of short-range correlations should
 contain some new physical information, and are not a simple result of the pseudorapidity distribution of
 final particles. The AMPT results with and without hadronic scattering are
compared. It is found that hadron scattering can only increase the
short-range correlations to some level, but is not responsible for
the different correlation shapes for different energies. Further
study shows that the different pseudorapidity dependence of
short-range correlations are mainly due to partonic evolution and
the following hadronization scheme.
\end{abstract}
 

 \maketitle
\clearpage
\section{Introduction}
The main goal of high energy heavy ion collision experiments is to
investigate the nuclear matter properties under extreme conditions
and detect quark-gluon plasma (QGP). It has been confirmed that QGP
is produced in relativistic heavy ion collisions~\cite{QGP1, QGP2}.
Many observables have been used to study and compare the properties
of nuclear matter with and without QGP, such as the strong
anisotropic collective flow~\cite{elliptic-flow}, nuclear
modification factor $R_{AA}$~\cite{Raa} and so on.

If a shortening in the correlation length in pseudorapidity space is
observed, it may be taken as a signal of transition to quark-gluon
plasma~\cite{transition2}. Therefore, particle correlations are also
of particular importance for the study of QGP properties. It has
been reported that inclusive two-particle correlations have two
components: short-range correlations (SRC) and long-range
correlations (LRC), corresponding to small and large pseudorapidity
differences, respectively.

Forward-backward correlations, as a special case of long-range
correlations, have been extensively studied in elementary particle
collisions~\cite{LRC-pp}~\cite{LRC-review} and by the STAR
experiment in heavy ion collisions at RHIC~\cite{LRC-STAR}. Because
they probe the longitudinal characteristics of the system produced
in heavy ion collisions, they provide an insight into space-time
dynamics and the earliest stages of particle production.

In comparison to the well-investigated long-range correlations that
determine the bulk properties of the collision system, short-range
correlations have a more subtle influence on the in-medium
properties of the particles~\cite{PHD-2003}~\cite{PHD-2007}. It is
believed that in pp collisions SRC arise due to the tendency of the
secondary particles to be grouped in clusters, which finally decay
to the real physical hadrons~\cite{SRC-pp1}~\cite{SRC-pp2}. To our
knowledge, few studies have been done about SRC in Au+Au collisions.

In this study, we will focus on the pseudorapidity dependence of
short-range correlations in Au+Au collisions. We will utilize a
multi-phase transport (AMPT) model ~\cite{ampt} to simulate Au+Au
collisions at RHIC, and try to record a complete system evolution.
The pseudorapidity dependence of short-range correlations for the
following cases are measured: (1) the effects of initial state;
(2)the early partonic phase; (3)the intermediate hadronic phase,the
time when the collision system undergoes the hadronization, while
hadronic scattering scheme is closed; (4)the hadronic phase, when
hadronization has happened and the hadronic scattering scheme is
open. These can present a clear image of short-range correlations
for different states of the collision system, especially for the
partonic phase and hadronic phase.

This paper is organized as follows. A short introduction of the AMPT
model is given in Section II. In Section III, we present the AMPT
results for the pseudorapidity dependence of neighboring-bin
correlation patterns in Au+Au collisions at $\sqrt{s_{NN}} $=
$7-62.4$GeV. We discuss the effects of both the partonic and
hadronic evolutions, and take $\sqrt{s_{NN}} $= 7.7GeV and 62.4GeV
as two examples to focus on the influence of parton phase evolution
on the short range pseudorapidity correlation patterns. Finally,
some conclusions are given in Section IV.

\section{A brief introduction to the AMPT model}
The Monte Carlo event generator AMPT (A Multi-Phase Transport) has
been used in this study. The AMPT model is made up of four main
components: initial conditions, partonic interactions, hadronization
and hadronic interactions. The initial conditions, which include the
spatial and momentum distributions of minijet partons from hard
processes and strings from soft processes, are obtained from the
Heavy Ion Jet Interaction Generator (HIJING) model~\cite{HIJING}.
The evolution  of partonic phase is modeled by Zhang's Parton
Cascade (ZPC) ~\cite{ZPC},  which includes only parton-parton
elastic scatterings with cross sections obtained from the pQCD
calculation with screening masses. The AMPT model has two versions,
the default AMPT model and the AMPT model with string melting
mechanism. In the default AMPT model, only minijet partons from the
initial conditions take part in the interactions modeled by ZPC.
When stoping interactions, they are combined with their parent
strings to form new strings. The resulting strings are then
converted to hadrons according to a Lund string fragmentation
model~\cite{fragmentation1, fragmentation2}. In the AMPT model with
string melting, the parent strings first fragment into partons and
then enter the ZPC model together with the minijet partons. After
freezing out,  a simple quark coalescence model is used to combine
the two nearest partons into a meson and three nearest partons into
a baryon~\cite{coalescence}. For both versions of the AMPT model,
the interactions among resulting hadrons are described by a
relativistic transport (ART) model~\cite{ART}.

Compared with the default AMPT model, the partonic phase can be
better modeled by the AMPT model with string melting mechanism.
Because the QGP is expected to be formed in heavy-ion collisions at
the Relativistic Heavy-Ion Collider(RHIC),  the AMPT with string
melting mechanism is considered a more efficient tool to study the
properties of the new matter, e.g. studying the elliptic flow and
triangular flow~\cite{flow}. Based on this, we will utilize the AMPT
model with string melting mechanism to generate Au+Au collisions at
$\sqrt{s_{NN}} = 7.7-62.4$ GeV. The parton cross section is taken to
be 10 mb in our simulations.

\begin{figure}
\begin{center}
\includegraphics[scale=0.35]{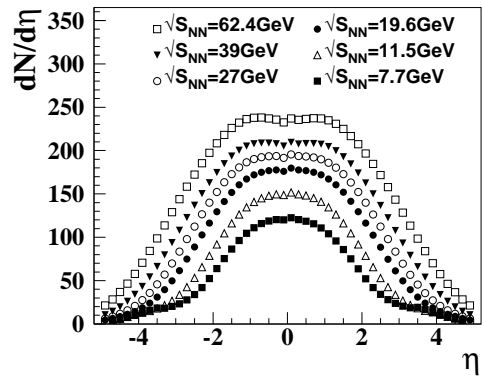}
\caption{\label{Fig1} The pesudorapidity distribution of final
particles at $\sqrt{s_{NN}}$ =$7.7-62.4$GeV Au+Au collisions using
the AMPT model with string melting mechanism. }
\end{center}
\end{figure}

\begin{figure*}
\includegraphics[width=6.2in]{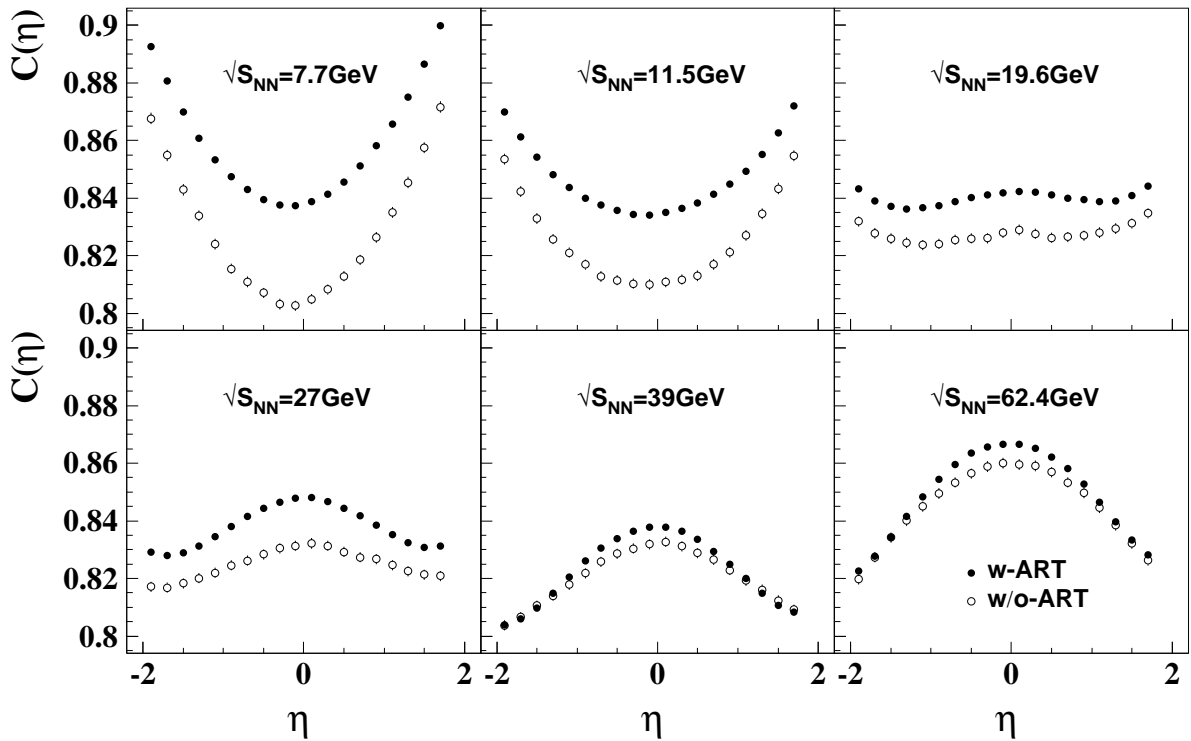}
\caption{\label{Fig2} The pseudorapidity dependence of neighboring
bin correlation pattern at  $\sqrt{s_{NN}}$ =$7.7-62.4$GeV Au+Au
collisions using the AMPT model with string melting mechanism, where
the solid points and open points correspond to the result with and
without hadronic scattering respectively. }
\end{figure*}

\section{Neighboring rapidity bin multiplicity correlation }
One focus of the analysis of the final state is on the longitudinal
momentum distributions, and correlations. In this section,  we show
the variation in particle density with $\eta$. Then, we mainly study
the pseudorapidity dependence of neighboring bin multiplicity at
different colliding energies.

Figure~\ref{Fig1} shows the pseudorapidity distributions of final
particles within -5 $\le$ $\eta$ $\le$ 5 in Au+Au collisions at
$\sqrt{s_{NN}} $= $7-62.4$GeV.  As expected the particle density
increases with decrease in $\eta$ for all energies. In addition, as
a function of collision energy, we can see the pseudorapidity
distribution grows systematically both in height and width.

The general two-bin multiplicity correlation is defined as
\begin{equation} C_{1,2}=\frac{\langle n_{1}n_{2}\rangle}{\langle
n_{1}\rangle\langle n_{2}\rangle}-1, \end{equation} where $n_{1}$
and $n_{2}$ are the numbers of particles measured in bin $1$ and bin
$2$. If particles are produced independently over the whole region,
then$\langle n_{1}n_{2}\rangle=\langle n_{1}\rangle\langle
n_{2}\rangle$ and correlations vanish.

It has previously been demonstrated that two specific 2-dimensional
correlation pattern, i.e. fixed-to-arbitrary and neighboring bin
correlation patterns, are efficient for identifying various random
multiplicative cascade processes~\cite{wu-pre}.

In this work, we apply the neighboring bin pseudorapidity
correlation pattern to the relativistic heavy-ion collisions,
proposing the following form
\begin{equation} C( \eta)=\frac{\langle n_{\eta}n_{\eta+\delta
\eta}\rangle}{\langle n_{\eta}\rangle\langle n_{\eta+\delta
\eta}\rangle}-1, \end{equation} where  $\delta \eta$ corresponds to
the width of the bins,  $\eta$ and $\eta+\delta\eta$ are the center
of the two bins, and $n_{\eta}$ and $n_{\eta+\delta \eta}$
correspond to the multiplicities of the two bins. In the following,
the mid-rapidity [-2, +2] is chosen and we divide the region equally
into 20 bins, which corresponds to the bin width  $\delta \eta
=0.2$. It is clear that the neighboring bin multiplicity correlation
pattern can demonstrate how the short range correlations vary with
the pseudorapidity.

\begin{figure*}
\includegraphics[width=6.2in]{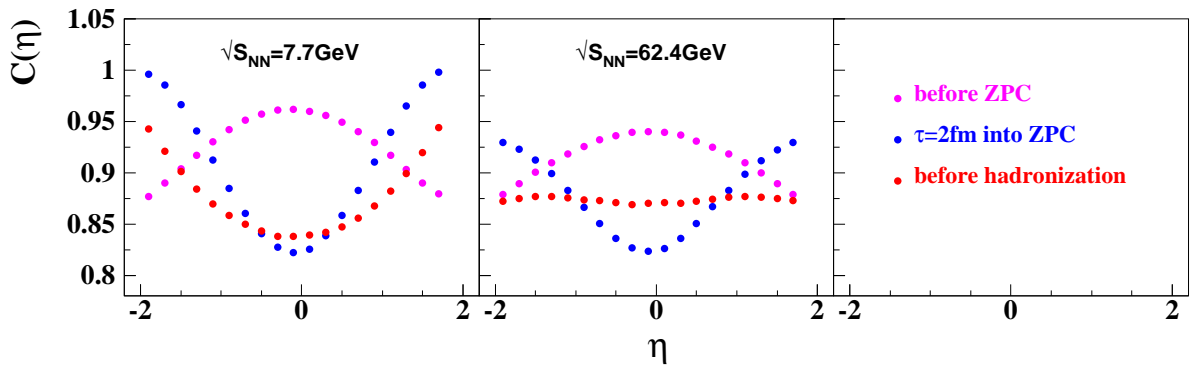}
\caption{\label{Fig3} The pseudorapidity dependence of neighboring
bin correlation pattern for partons at  $\sqrt{s_{NN}} $= 7.7GeV and
62.4GeV Au+Au collisions using the AMPT model with string melting
mechanism, where different colors correspond to the correlation
results from different evolution times. }
\end{figure*}

Figure~\ref{Fig2} presents the AMPT results for the dependence of
neighboring bin correlation patterns for Au+Au collisions at
$\sqrt{s_{NN}} $= 7.7, 11.5, 19.6, 27, 39 and 62.4 GeV. We calculate
two cases of correlations, i.e. (a) ¡°w-ART¡±, which represents the
results from a complete time evolution process of the AMPT model, as
shown by solid circles; (b) ¡°w/o- ART¡± represents the results from
the time just after quark coalescence, but before hadronic
scatterings, shown as open circles. Comparing these two cases, can
help us understand the hadronic effect on the observable. From the
figure, we can see the neighboring bin correlation patterns have a
different dependence on the pseudorapidity at different colliding
energies. We can see that both ¡°w-ART¡± and ¡°w/o-ART¡± show that
the correlation strength first decreases and then increases with the
pseudorapidity, and has a minimum value at the central
pseudorapidity for Au+Au collisions at $\sqrt{s_{NN}}$ = 7.7 and
11.5GeV, while it first increases and then decreases, with a maximum
at the central pseudorapidity, for Au+Au collisions at
$\sqrt{s_{NN}}$ = 27, 39 and 62.4 GeV. For Au+Au collisions at
$\sqrt{s_{NN}}$ = 19.6GeV, the correlation patterns have little
dependence on the pseudorapidity. It is well known that the
pseudorapidity distribution has a plateau at mid-rapidity for all
energies, as shown in Figure~\ref{Fig1}, thus we believe that the
pseudorapidity dependence of short range correlations should contain
some new physical information, but is not a simple result of the
pseudorapidity distribution of particles. From the figure, we can
see the results from ¡°w/o-ART¡± are lower than that those ¡°w-ART¡±
for all collision energies, which means hadronic scattering can
increase the correlation strength to some level.In addition, the
values of ¡°w/o-ART¡± are closer to those of ¡°w-ART¡± at higher
energies. This indicates that hadronic interactions have little
effect on the correlation patterns at very high energies.

As we know, the AMPT model is a hybrid model, in which hadronic and
partonic interactions are both included~\cite{ampt}. It is widely
believed that hadronic interactions dominate the correlation
behaviors at low energies and partonic interactions dominate the
correlation behaviors at high energies~\cite{Nu-QCD}. In
Figure~\ref{Fig2}, we learn that there exist different correlation
behaviors for different energies, and that hadronic interactions
only have an effect on the correlation strength, cannot explain the
different pseudorapidity dependence of short-range correlations with
collision energy. Further study is needed.

Next, Au+Au collisions at $\sqrt {s_{NN}}$ = 7.7GeV and 62.4GeV are
chosen to study the effect of the partonic phase on the correlation
behavior, since they correspond to the two cases for low and high
energy respectively. In Figure~\ref{Fig3},  we present the
pseudorapidity distribution of short-range correlations between
partons for the two energies at different partonic evolution times.
Three important evolution times are considered. The ¡°before ZPC¡±
represents the short range correlation pattern from the initial
state of partonic matter; ¡°$\tau=2 fm/c$ into ZPC¡± describes the
short-range correlation pattern at the time point when the partonic
phase evolution has been going on for 2 $fm/c$. The ¡°before
hadronization¡± presents the short-range correlation pattern from
the moment when all of partons freeze out before hadronization.

From Figure~\ref{Fig3}, we can see that the short-range correlations
first increase and then decrease with pseudorapidity both for 7.7GeV
and 62.4GeV for the initial state of the colliding system (i.e.
¡°before ZPC¡±). For ¡°$\tau=2 fm/c$ into ZPC¡±, it is interesting
to note that the correlation patterns show the reversal effect, as
they first decrease and then increase with pseudorapidity for both
energies. When the partons stop interactions (¡±before
hadronization¡°), however, there is a significant difference for
Au+Au collisions at 7.7 GeV and 62.4 GeV. For 7.7GeV Au+Au
collisions, the short-range correlations continue to keep a curved
upward trend with pseudorapidity, while for 62.4GeV Au+Au
collisions, the short-range correlations are not sensitive to the
pseudorapidity. This can be explained by the fact that for $\sqrt
{s_{NN}}$ = 62.4GeV Au+Au collisions there exist enough interactions
between partons to drive the colliding system into isotropy.

After partons in the string melting scenario stop interacting, their
hadronization is modeled via a simple quark
coalescence\cite{coalescence}. In the quark coalescence model, we
can combine the two nearest partons into a meson and three nearest
quarks into a baryon. Related research\cite{yield-parton} has shown
that the total yield of the meson (baryon) is proportional to the
square (cube) of quark density just before hadronization. We argue
that correlation of final particles is also related to the
constituent quark density in the hadronization process. By combining
the nearest quarks into hadrons, the hadronization scheme of quark
coalescence can obviously increase short-range correlations, and the
influence of the hadronization process on short-range correlations
strongly depends on the quark density just before hadronization. For
$\sqrt {s_{NN}}$ = 62.4GeV Au+Au collisions, the flat correlation
pattern as shown in Figure~\ref{Fig3} become a shape curved downward
after hadronization, due to stronger short-range correlations from
the larger quark density at central pseudorapidity just before
hadronization. However, because only a limited number of partons
pass through hadronization for 7.7 GeV Au+Au collisions, the weak
influence doesn't change the correlation shape and it keep a shape
curved upward. By combining the effects of partonic evolution and
the subsequent quark coalescence scheme, the different
pseudorapidity dependence of neighboring bin correlation pattens at
different energies, as shown in Figure~\ref{Fig2}, is
understandable.

\section{Conclusions}
In summary, the pseudorapidity dependence of neighboring bin
multiplicity correlation patterns are studied in a multi-phase
transport model with both partonic and hadronic interactions. It is
shown that the short-range correlations vary significantly with
pseudorapidity for different energy regions. In particular, for
$\sqrt{s_{NN} } < $19.6GeV Au+Au collisions,  the short-range
correlations have a trough at central pseudorapidity, while for
$\sqrt{s_{NN}} > $19.6GeV Au+Au collisions, the short-range
correlations of final particles have a peak at central
pseudorapidity. The AMPT results with and without hadronic
scattering are compared. It is found that hadron scattering can only
increase the short-range correlations to some level, but is not
responsible for the different correlation shapes for different
energies. Further study shows the differences in pseudorapidity
dependence of short-range correlations are mainly due to partonic
evolution and the following quark coalescence mechanism. We argue
that the neighboring bin multiplicity pseudorapidity correlation
patterns are sensitive to the dynamics mechanism of the collision
system evolution, and especially have a different response to the
partonic phase and hadronic phase.

\section{Acknowledgments}
The author Wang Meijuan thanks Xu Mingmei and Zhou You for useful discussion. 
This work were supported by GBL31512, Major State Basic 
Research Devolopment Program of China under Grant No. 2014CB845402, 
the NSFC of China under Grants  No.11475149, No. 11175232, 
No. 11375251, No. 11421505 and  No. 11221504.

\end{document}